\documentclass[aps,prl,twocolumn,superscriptaddress,showpacs]{revtex4-1}
\usepackage{graphicx}
\usepackage{MnSymbol}

\begin{document}

%Title of paper
\title{Phonons in potassium doped graphene: the effects of electron-phonon interactions, dimensionality and ad-atom ordering}

\author{C. A. Howard}
\email{c.howard@ucl.ac.uk}
\affiliation{London Centre for Nanotechnology, University College London, London WC1E 6BT, UK}
\affiliation{Department of Physics, Royal Holloway, University of London, Egham, Surrey, TW20 0EX, UK}
\author{M. P. M. Dean}
\email{mdean@bnl.gov}
 \affiliation{Department of Condensed Matter Physics and Materials Science,Brookhaven National Laboratory, Upton, New York 11973, USA}
\affiliation{Cavendish Laboratory, University of Cambridge, JJ Thomson Avenue, Cambridge CB3 0HE, UK}
\author{F. Withers}
\affiliation{Centre for Graphene Science, School of Physics, University of Exeter, Exeter, EX4 4QL, UK}

% user macros
\newcommand{\microns}{\ensuremath{\mathrm{\mu}}m}
\newcommand{\degc}{\ensuremath{^{\circ}}C}
\newcommand{\invcm}{cm$^{-1}$}
\newcommand{\e}[1]{$\times$10$^{#1}$}
\def\mathbi#1{\textbf{\em #1}}

\date{\today}

\begin{abstract}

Graphene phonons are measured as a function of electron doping via the addition of potassium adatoms. In the low doping regime, the in-plane carbon G-peak hardens and narrows with increasing doping, analogous to the trend seen in graphene doped via the field-effect. At high dopings, beyond those accessible by the field-effect, the G-peak strongly softens and broadens. This is interpreted as a dynamic, non-adiabatic renormalization of the phonon self-energy. At dopings between the light and heavily doped regimes, we find a robust inhomogeneous phase where the potassium coverage is segregated into regions of high and low density. The phonon energies, linewidths and tunability are remarkably similar for 1-4 layer graphene, but significantly different to doped bulk graphite.

\end{abstract}

% insert suggested PACS numbers in braces on next line
\pacs{63.22.Rc,78.67.Wj,81.05.ue}
% insert suggested keywords - APS authors don't need to do this
%\keywords{}

%\maketitle must follow title, authors, abstract, \pacs, and \keywords
\maketitle

%%%% INTRODUCTION
Due to the intense scientific interest in graphene over the past few years, many of its basic properties have been determined. Now much of the effort in graphene research is devoted to tuning its properties in order to search for exotic physics and to extend and improve its potential for applications \cite{Geim2009,CastroNeto2009}. The properties of graphene can be tuned both by varying the number of layers in the graphene stack and via doping \cite{CastroNeto2009, Pisana2007,Yan2007,Jung2009,Zhao2011,Alzina2010,Bruna2011, Das2008, Chen2011}.
The current method of choice for doping graphene is via the electric field effect \cite{Pisana2007,Yan2007}. In this way the Fermi level can be controllably tuned to a maximum of $E_{D}=$-0.3~eV away from the Dirac point (about 0.002~e$^-$/C atom) giving carrier densities of $\sim$10$^{13}$~cm$^{-2}$. Similar levels of doping have also been achieved via the addition of Br$_2$ \cite{Jung2009}, FeCl$_3$ \cite{Jung2009, Zhao2011}, O$_3$ \cite{Alzina2010} and CHF$_3$ \cite{Bruna2011} and higher values ($E_D \approx$0.8~eV) can be obtained using electrolytic gating \cite{Das2008, Chen2011, Ye2011}. The deposition of alkali metal atoms provides a route to even greater doping levels and in this way the Fermi level can be incrementally moved to $E_D$=-1.3~eV (0.03~e$^{-}$/C atom or $\sim$10$^{14}$~cm$^{-2}$) \cite{Bostwick2007,Bianchi2010}.

As the electronic structure is modified, so too is the electron-phonon interaction (EPI) \cite{Pisana2007,Yan2007,Das2008,Lazzeri2006,Attaccalite2010}. A detailed understanding of this interaction is of great importance as it not only governs electronic transport, and hence the performance of graphene based electronic devices, but can also mediate exotic ground states such as superconductivity and charge density waves. At light doping levels a small (0.3\%) hardening in the in-plane carbon phonon energies and narrowing in their linewidth have been reported \cite{Pisana2007, Yan2007,Jung2009,Zhao2011,Alzina2010,Bruna2011, Das2008, Chen2011}. This is due to a \emph{reduction} in the electron-phonon scattering as the Kohn anomaly found in pure graphene at $\boldsymbol\Gamma$ is gradually removed to finite \mathbi{q} \cite{Pisana2007, Yan2007}. Here we extend the investigation of graphene phonons to higher dopings where we discover both a strong (3\%) softening and significant linewidth broadening of the in-plane carbon phonons. We argue these effects are due to a novel, dynamic EPI arising from the 2D metallic nature of heavily doped graphene. In addition, we find that the tunability, phonons and EPI are remarkably similar for 1-4 layer doped graphene, but these systems exhibit significantly different behavior to doped bulk graphite.

%%%% EXPERIMENTAL METHODS
Graphene was prepared by micromechanical exfoliation of natural graphite onto an oxidized Si substrate (275 nm SiO$_2$) \cite{suppmat}. The substrate was then loaded into a sealed borosilicate tube with an optical window, evacuated and outgassed at 250\degc{} for 24 hours. An ingot of potassium metal was then added in a high purity argon glovebox, the tube was evacuated and then introduced into a furnace. The level of doping was incrementally increased by repeatedly exposing the graphene to the potassium vapor. The bulk potassium graphite intercalation compounds (GICs), KC$_8$ and KC$_{24}$ were made by the vapor transport method \cite{Dresselhaus2002}. Raman experiments were performed using a Renishaw inVia micro-Raman Spectrometer equipped with a 514.5~nm laser. The laser was focused to $\sim$3~\microns{} and the power at the sample was kept below 2~mW.

\begin{figure}
\includegraphics{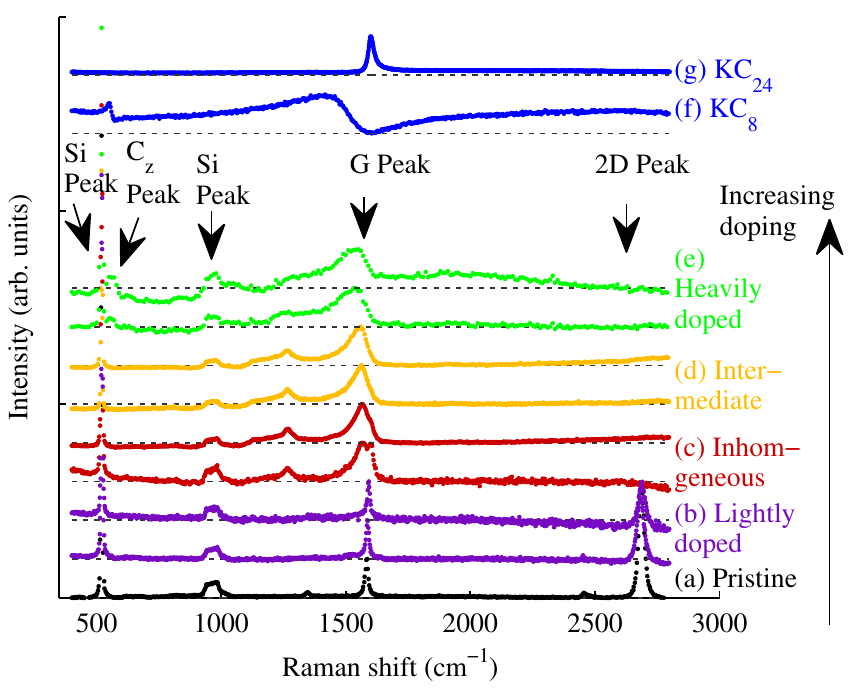} %
\caption{(color online). The Raman spectra of graphene in order of increasing potassium doping. The approximate positions of the peaks are marked by arrows. Representative spectra are offset from one-another and normalized to the G-peak for clarity. The different doping regimes are denoted to the right of the plot: (a) undoped (b) lightly doped (c) inhomogeneous (d) intermediate (e) heavily doped. Also shown are the Raman spectra for the GICs (f) KC$_8$ and (g) KC$_{24}$\label{Fig1}}
\end{figure}

Figure~\ref{Fig1} shows the evolution of the Raman spectra of graphene in order of increasing K doping. The development of the features allow us to identify five main doping regimes: pristine (undoped), lightly, inhomogeneous, intermediate and heavily doped graphene. For comparison, we plot Raman spectra of KC$_8$ and KC$_{24}$. The Raman spectrum of pristine graphene Fig.~\ref{Fig1}(a) is well known \cite{Ferrari2006}. The peak at 1583~\invcm{} is an E$_{2g}$ symmetry phonon at $\mathbf{\Gamma}$ and is commonly termed the G-peak. The peak at 1350~\invcm{} is the D-peak which arises from an in-plane transverse optical phonon around the \textbf{K} point in the Brillouin zone and is activated by disorder scattering \cite{Thomsen2000}. The intense single component peak at 2686~\invcm{} is the second order relative of the D peak and is a fingerprint of monolayer graphene. The spectra of lightly doped graphene (Fig.~\ref{Fig1}(b)) are qualitatively similar to pristine graphene, here the G-peak is sharp and single component indicating homogeneous doping. Upon further doping the G-peak is split into two components and the 2D peak disappears (Fig.~\ref{Fig1}(c)). The splitting of the G-peak indicates inhomogeneous doping/K coverage and is discussed in more detail below. The disappearance of the 2D peak could be associated with a removal of the resonance conditions by the raised Fermi level (i.e.\ when the energy of the incident light, $E_L<2E_D$), however this is unlikely at this level of doping given the large laser energy (2.41~eV). Furthermore we also find an absence of a 2D peak in bulk KC$_{24}$ (Fig.~\ref{Fig1}(g)) despite the Dirac point in this material being measured to be -0.75~eV \cite{Pan2011}, well within the resonance condition. Theoretical calculations predict the suppression of the 2D peak intensity with doping \cite{Basko2009}, but our measurements indicate the suppression of this peak is somewhat faster than predicted. Thus our work questions the validity of using the disappearance of this peak to determine the doping level in graphene.

In the intermediate regime (Fig.~\ref{Fig1}(d)), a single-component G-peak is recovered which is downshifted and broadened. Finally, at the highest dopings (Fig.~\ref{Fig1}(e)), the G-peak is accompanied by the appearance of another Raman mode at 560~\invcm{}. This mode coincides in energy with a mode in KC$_8$ (Fig.~\ref{Fig1}(f)). This compound consists of stacked graphene sheets separated by potassium layers \cite{Dresselhaus2002}. The mode exists at the \textbf{M} point of the graphene Brillouin zone but is folded to $\mathbf{\Gamma}$ by the 2$\times$2 larger in-plane unit cell and becomes Raman active. Thus the presence of this mode indicates the regions of a 2$\times$2 ordered potassium lattice on the graphene. As this mode involves motion of carbon atoms perpendicular to the graphene planes we term it the C$_z$ peak. The relative intensity of the C$_z$ peak increases with increasing doping whilst the G-peak continues to soften and broaden until the spectra no longer changes with further K exposure.

For all dopings higher than lightly doped graphene, additional modes appear in the region 1100-1300~\invcm{}. The origin of these features is unclear. Although these may be related to the graphene D-peak, they exist up to saturation doping where the Dirac point is measured to be -1.29~eV \cite{Bianchi2010}. Here, the resonant mechanism is forbidden and the D-peak would be expected to have negligible intensity. Another explanation is that these features are $\mathbf{\Gamma}$ point phonons that are Raman inactive but become visible due to disorder of the potassium atoms on the surface. These features will be discussed in more detail elsewhere \footnote{C.~A.~Howard, M.~P.~M.~Dean, and F.~Withers in preparation}.

The G-peak shows a strong change in character with doping. To investigate this in more detail, this feature is fitted with the asymmetric Breit-Wigner-Fano lineshape. This lineshape is due to coupling between the phonon and an electronic continuum \cite{Fano1961}, and it is commonly found in the Raman spectra of doped graphitic systems. It is modelled as a signal of intensity:
\begin{equation}
I(\omega)=I_0 \frac{(1+\frac{\omega-\omega_{ph}}{q\Gamma/2})^2} {1+(\frac{\omega-\omega_{ph}}{\Gamma/2})^2} \label{BWFeqn}.
\end{equation}
Here $1/q$ quantifies the asymmetry of the shape and $\omega_{ph}$ and $\Gamma$ are fitting parameters to the central frequency and full width at half maximum (FWHM) of the bare phonon. Example fits and a detailed discussion of the fitting efficacy, background and Fano resonance are given in \cite{suppmat}. We find a $1/q$ of -0.2 --- -0.3, and that this parameter shows no clear trends with doping.

\begin{figure}
\includegraphics{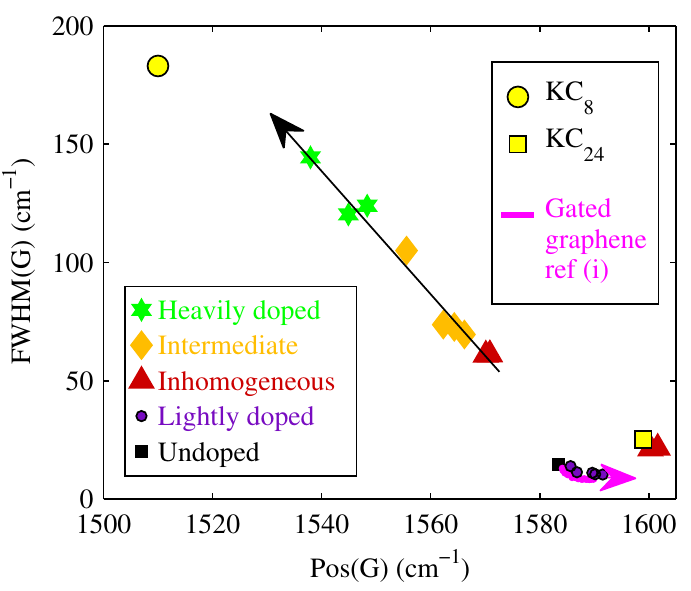}
\caption{(color online). The correlation between the G-peak energy and width for the different doping regimes: undoped (black  $\filledsquare$), lightly doped (purple $\bullet$), inhomogeneous (red $\filledmedtriangleup$), intermediate (yellow$\filleddiamond$), and heavily doped (green $\filledstar$). These are  compared with results on gated graphene from \cite{Casiraghi2009,Pisana2007} (magenta line). The arrows are guides to the eye depicting the trend with increasing doping. \label{Fig2}}
\end{figure}

Fig.~\ref{Fig2} shows the change in the width and energy of the G-peak as a function of increasing doping, where we also compare our results to gated graphene \cite{Casiraghi2009,Pisana2007}, KC$_8$ and KC$_{24}$. For lightly K-doped graphene the G-peak hardens and narrows closely following the trends found in gated graphene \cite{Pisana2007,Yan2007, Das2008}. This is well understood: in undoped graphene, there is a Kohn anomaly at $\mathbf{\Gamma}$ which softens the G-peak and increases its linewidth \cite{Lazzeri2006}. As the graphene is lightly doped the Kohn anomaly is gradually shifted to finite \mathbi{q}, where it no-longer interacts with the Raman phonons at \mathbi{q}$\sim$ 0. \cite{Pisana2007,Yan2007,Das2008} Comparison with data for gated graphene allows us to estimate the maximum doping in this region to give $E_{D} \approx $-0.3~eV \cite{Das2008}. At heavier dopings there is an abrupt crossover in behavior and the G-peak significantly softens. We propose this change in energy is due to the charge transfer into the antibonding $\pi^*$ electronic bands. This destabilizes the carbon-carbon bonds and thus softens the phonon. Similar behavior is found in GICs where a measured increase in bond length has been correlated with an increase in electron doping \cite{Dresselhaus2002, Pietronero1981,Dean2010Raman,Dean2010neutrons}.

The softening is accompanied by a large broadening of the linewidth indicative of a reduction of the phonon lifetimes with increasing doping. This is unlikely to arise from disorder: we measure an even greater width in an ordered bulk crystal of KC$_8$ of 185~\invcm{}. Anharmonic effects in graphitic systems are also typically far smaller than the linewidths reported here \cite{Dean2010Raman, Saitta2008}. EPIs are therefore the most likely cause of the decreased phonon lifetimes. 

Engelsberg and Schrieffer \cite{Engelsberg1963} were the first to predict that in certain metals, when the electron scattering rate, becomes comparable to or slower than the phonon frequency the resulting \emph{dynamic} EPI can result in a significant non-adiabatic renormalization of the phonon self energies. Here the normally small ($\sim$1\%) correction to a \emph{static} consideration of the EPI can become far larger provided the condition $|\textbf{\em{q.v}}_F| \ll \omega$, \cite{Engelsberg1963,Maksimov2008, Saitta2008} is fulfilled. Here $\mathbi{q}$ is the phonon wavevector, $\textbf{\em{v}}_F$ is the Fermi velocity and $\omega$ is the phonon frequency. At the same time the system must be a good metal with a significant density of states at the Fermi level. Consequently, these effects are most important for low dimensional metals, when $\mathbi{q}$ is parallel to a direction in which $\textbf{\em{v}}_F$ is small. Recent work has shown that this mechanism provides a justification for the large linewidths found in bulk GICs and MgB$_2$ \cite{Saitta2008,Dean2010Raman, Calandra2010}. However, given the non-tunability of these materials, monitoring this novel EPI as a function of increasing charge carriers has not been possible until now. To this end doped graphene, a tunable 2D metal, presents the idealized system to realize and investigate these effects. As the doping is incremented the 2D $\pi^*$ bands are populated, increasing the phase space for the EPI and therefore decreasing the phonon lifetimes. Thus these large linewidths are consistent with a large dynamic EPI. The two distinct trends of the G-peak with low and high doping shown in Fig.~\ref{Fig2}, highlight the different physical process involved: at low doping the phonon energies and lifetimes are dominated by the Kohn anomaly whilst at higher doping charge transfer and the effects of a large dynamic EPI. 

\begin{figure}
\includegraphics{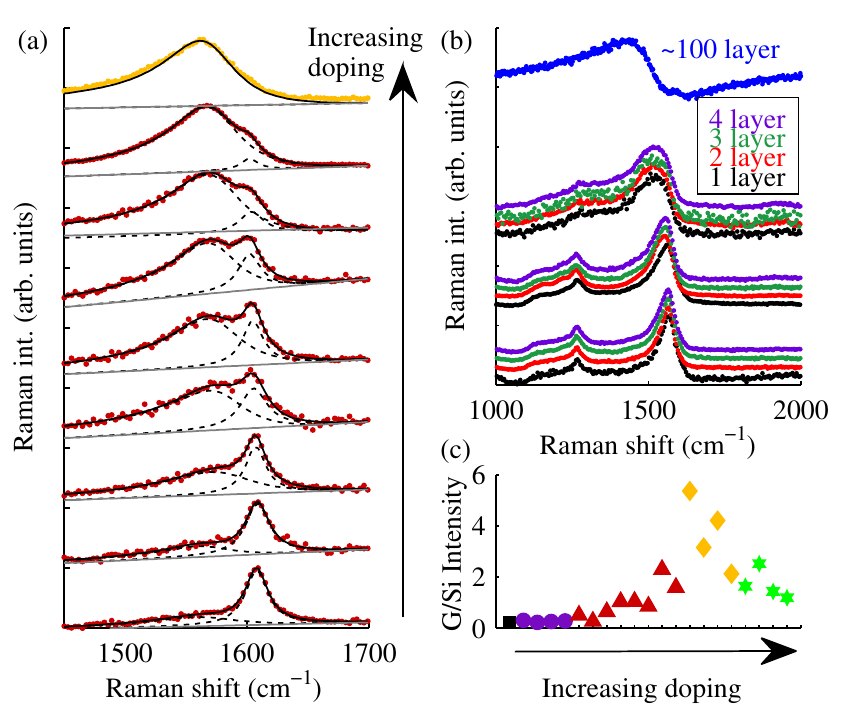} %
\caption{(color online). a) The evolution of the G-peak of inhomogeneously doped graphene with doping b) The G-peak of 1-4 layer graphene for three dopings including saturation, and the G-peak of a saturation doped $\sim$100 layer graphite flake c) The effect of doping the G-peak integrated intensity (normalized to the 520~\invcm{} Si peak), labels are as in Fig 2.\label{Fig3}} 
\end{figure}

All attempts to investigate the crossover between the two trends in Fig.~\ref{Fig2} resulted in the formation of the inhomogeneous phase which was found to be a robust intrinsic phase visible over a range of low K dopings. The development of this phase is plotted in more detail in Fig~\ref{Fig3}(a). The higher energy peak is comparable in energy to lightly doped graphene; the lower energy peak to a more heavily doped region. As the doping is incremented the intensity of the lower energy peak increases as the higher energy peak decreases, consistent with the two peaks arising from two distinct phases. The higher energy peak is of very similar width and energy to that of bulk KC$_{24}$ (Fig.~\ref{Fig1}(g)). This compound has a stable, homogeneously dispersed coverage of K atoms, which we propose forms in K-graphene and is maintained by the electrostatic repulsion of K ions. As the doping is increased further, K atoms are accommodated in distinct highly doped regions which increase to eventually cover the entire sample. We found no variation in relative peak heights as we moved the beam around on the sample indicating that the domain regions are much smaller than the laser spot size ($\sim$3~\microns{}). 

In order to further explore the effect of dimensionality we incrementally doped 1-4 layer graphene on the same substrate so the exposure to potassium vapor was identical for each sample. This data is shown in Fig.~\ref{Fig3}(b) where we also compare our results to bulk KC$_8$. Remarkably, we found at five different dopings, within error of 4~\invcm{}, the width and energy of the G-peak are independent of the number of layers. We can therefore conclude there is no doping below the graphene sheets which would result in higher average charge transfer for the mono-graphene compared with 4 layer graphene and, that the tunability, charge transfer and EPI interactions are very similar for 1-4 layer graphene. These results indicate that potassium doped Few Layer Graphene (FLG) behave like a stack of non-interacting decoupled monolayers. This is significant because the detailed electronic structure of FLG differs depending on the number of layers and their stacking \cite{CastroNeto2009}. Upon intercalation with K, the increased separation of the graphene sheets and their expected restacking from A/B to A/A sequence as found in KC$_8$ \cite{Rudorff1954}, account for the similarities in the behaviour of K-doped FLG. We found a crossover from the 2D K-graphene spectra to bulk spectra in a thin graphite flake of $\sim$100 layers (Fig.~\ref{Fig3}(b)) \cite{suppmat}.  

It is notable that the G-peak of bulk KC$_8$ has an even greater width and lower energy than saturated K-graphene. This is consistent with the fact the maximum doping achieved is lower in K-graphene, where $E_D=-1.29$~eV \cite{Bianchi2010} than in KC$_8$ \cite{Gruneis2009a, Pan2011} ($E_D=-1.35$~eV). We have shown that exposing 1-4 layer graphene to K vapor permits a tunable increase in doping, rather than the distinct stoichiometric compounds formed when bulk graphite is treated in the same way \cite{Dresselhaus2002}. We also found a large difference in the kinetics of the doping: for the same time to form lightly doped graphene, KC$_8$ would form from a bulk graphite flake on the same substrate. These contrasting behaviors can be explained by the crucial difference between K-graphene and K-GICs: the lack of long range interlayer interactions in the former system. For example, bulk KC$_8$ forms a unit cell with K atoms correlated over 4 graphene layers (21.4~\AA{}) \cite{Rudorff1954}. Whilst the lack of these interactions in K-graphene results in this systems tunability, it is likely that this also inhibits the complete K coverage, limiting the doping achievable, and introducing intrinsic disorder into the adlayer.

Fig.~\ref{Fig3}(c) shows a maximum in the normalized integrated intensity of the G-peak at intermediate doping. A similar trend has been seen as graphene is electrostatically hole-doped \cite{Chen2011}. In this work, the authors show that the blocking of the resonant production of electron-holes, i.e. when $E_L<2E_D$, causes an \emph{increase} in the G-peak intensity as the destructive quantum interference existing between the different inelastic pathways is reduced \cite{Chen2011}. Our results confirm this effect for electron doped graphene and allow us to identify the doping to give E$_D$=-1.2~eV at the maximum in Fig.~\ref{Fig3}(c), if a simple analogy between this hole-doped gated structure and our potassium doping is valid. This is consistent with a maximum doping of about -1.3~eV.

In conclusion, we have shown that the rich evolution of the G-peak in graphene with doping presents a spectacular change of the EPI in this material. At low dopings the G-peak hardens and its linewidth decreases, analogous to trends found in gated graphene due to a decreasing EPI. In contrast, at high doping the G-peak significantly softens and broadens due to a large dynamic EPI. Unlike bulk graphite, we find that 1-4 layer graphene is tunable by exposure to potassium, important for tailoring the properties of graphene for applications. However, while at the high and low K dosings the doping is homogeneous, at in-between dosings segregated regions of high and low density K coverage co-exist. More generally, the diverse trends found in the tunable system of doped graphene provides a single system displaying the behavior found in all graphitic systems with doping, for example, explaining the contrasting linewidths found in the G-peaks of carbon nanotubes at light \cite{Bushmaker2009} and heavy doping \cite{Rao1997}.

\begin{acknowledgments}
We thank the EPSRC for funding, Felix Fernandez-Alonso, Andrew Walters and Mark Ellerby for fruitful discussions and Steve Firth for technical assistance. The work at Brookhaven is supported in part by the US DOE under contract No.\ DEAC02-98CH10886 and in part by the Center for Emergent Superconductivity, an Energy Frontier Research Center funded by the US DOE, Office of Basic Energy Sciences.
\end{acknowledgments}

\end{document}